\documentstyle[preprint,aps]{revtex}
\begin{document}
\draft
\title{The effects of the carrier interaction and electric fields\\
on subband structures of selectively--doped\\
semiconductor quantum wells}
\author{Sung--Kyun Park and Kyung--Soo Yi}

\address{Department of Physics, Pusan National University, Pusan
609--735, Korea}

\date{\today}
\maketitle

\begin{abstract}
\noindent
We investigate the ground--state electronic properties of the symmetrically--
doped semiconductor quantum well in the presence of a homogeneous electric
field. In this paper we examined the effect of the electric field and carrier
interaction on the subband structure as a function of the field strength
and carrier concentration.  The many--body effects are evaluated using a
local density functional exchange--correlation potential.
We find that the electron subband energy is reduced as the magnitude of the
electric field is increased, but it is increased as the surface carrier
density is increased.   However, the separation of the electron subband
energies is reduced for the increase in both the electric field and surface
carrier density.
On the other hand, the energy separation of the hole subbands
is increased as the carrier density and field strength are increased.
Effect of the exchange--correlation potential on the subband structure is
found negligibly small in this calculation. The subband energies are reduced
slightly, increasing their separations little in the presence of the local
exchange--correlation potential.
\end{abstract}
\vspace{0.5cm}

\section*{1. Introduction}

\noindent
In recent years extensive efforts have been devoted to investigating
the electronic properties of semiconductor quantum well
structures\cite{Bastard1}.
These studies were motivated by the improvements in the crystal growth
techniques, such as molecular beam epitaxy and metal organic chemical
vapor deposition. These growth methods are capable of producing
ultrathin layers of semiconductor heterostructures with sharp interfaces of
high quality.   A high--mobility semiconductor quantum well structures can be
realized with a use of the modulation--doping technique\cite{Dingle1}.
In modulation--doping technique, for example, only the AlGaAs layers are
doped with impurities and the GaAs layers are undoped.
Doped impurities (donor impurities in our case) in the AlGaAs layers are
ionized and electrons transfer to the GaAs layers.
In this case one needs to examine the subband structures self--consistently
taking into account the effects of charge transfer from AlGaAs to GaAs
layers and of the band bending.

In the presence of an external d. c. electric field, new structures and
phenomena were predicted and observed experimentally\cite{Miller1,Miller2}.
In particular, the tunability of the optical properties of the structure
in the presence of an external electric field were studied considerably in
detail\cite{Tsang,Wood}.   Far infrared intraband
absorption\cite{West,Levine1} was also observed in GaAs/AlGaAs layers with
high detectivity approaching that of HgCdTe\cite{Levine2}.
The quantum well high--electron--mobility transistor has also been
studied\cite{Mishra}.
Most of these studies involve some form of modulation of the quantum well
eigenenergies and eigenfunctions in the presence of an electric field for
their principal mode of operation.

Previous investigations have been focused in understanding the effects of the
electric field and carrier interaction separately employing methods of either
variational or perturbation calculations in a frame of the single-particle
picture. The variational method, although
computationally the simplest, has drawback of not knowing the accuracy and
region of validity\cite{Bastard2}.   Bloss investigated only the
electric field
effects by solving differential equation numerically and neglecting the
effects of carrier interaction\cite{Bloss}.
Ando investigated only the effect
of carrier interaction in the absence of the electric field\cite{Ando}.
However, in practical devices, one should take both of these two effects,
self--consistently, into account on an equal footing.

In this paper, we investigate the effects of the carrier interaction and
electric field on the electronic properties of symmetrically--doped
GaAs/AlGaAs single quantum well (SQW) structures. We evaluate the electron
and hole subband structures self--consistently within a Hartree approximation
in the presence of a d. c. external electric field perpendicular to the
interfaces. Also, a simplified local density functional exchange--correlation
potential is employed to include the many--body effects such as exchange and
correlation interactions in the calculation.
In section 2, we describe a self--consistent numerical calculation of subband
structures of the SQW in the presence of
a d. c. electric field and carrier intercation. In section 3, we present the
numerical results of GaAs/AlGaAs system.  The subband energies are presented
as a function of surface carrier density for several different values of the
field strength. Finally, we conclude the work in section 4.

\section*{2. Theory}

\noindent
We consider a symmetrically--doped GaAs/AlGaAs SQW structure in the presence
of an external d. c. electric field within the effective mass
approximation\cite{Bendaniel}.   Here we assume that the effective mass
encounters the effect of bulk band structure and we ignore the mismatch of
effective masses in the well and barrier regions.   The motion of carriers
(electrons or holes) in the $z$ direction, which is the growth direction of
the SQW and is perpendicular to the interface, is quantized due to the
one--dimensional confining potential well.  And, hence, the carriers possess
only two degrees of
freedom along the $x$ and $y$ directions.  Therefore, the single--particle
wave function describing the motion in the $xy$--plane is simply plane
wave like and the wave function in the $z$ direction is the solution of a
one--dimensional Schr\"{o}dinger equation

\begin{equation}
[-\frac{\hbar^2}{2m^*} \frac{{\rm d}^2}{{\rm d}z^2} + V_t (z)]
\psi(z) = E\psi(z).
                                            \label{schr}
\end{equation}

\noindent
Here $m^*$ is the effective mass of a carrier in the quantum well and
the effective potential energy function $V_t(z)$ subjected to a single
particle in the system is given by

\begin{equation}
V_t(z) = V_w(z) + V_F(z) + V_{s.c.}(z).
                                            \label{poten}
\end{equation}

\noindent
In equation\ (\ref{poten}) $V_w(z)$ is the rectangular well potential energy
associated with the
band--gap mismatch at heterointerfaces and $V_F(z)$ is the linear potential
energy due to the constant electric field applied along the growth direction.
And $V_{s.c.}(z)$ is the energy of a given carrier due to the interactions
with ionized impurities doped in the barrier material and with the
other remaining carriers in the quantum well.
The interaction potential energy
$V_{s.c.}(z)$ should be determined self--consistently.

If one simply replaces the wide band--gap material (AlGaAs layer) with a
simple potential barrier of height $V_o$, $V_w(z)$ is written by
[see figure 1]:

\begin{equation}
V_w (z) = \left \{ \begin{array}{ll}
                    0,   & |z| < L/2 \\
                    V_o, & |z| > L/2
                   \end{array}
          \right.
                                          \label{squar}
\end{equation}

\noindent
where $L$ is the width of the gap region (GaAs layer) i.e.,
the width of the quantum well.
In this work the band--gap mismatch $V_o$ is assumed to be
60\verb|%| (or 40\verb|%|) of the band--gap mismatch of AlGaAs and GaAs
for the
conduction (or valence) band\cite{Bloss}.
In the presence of an external d.c.
electric field $F$ in the $z$ direction, the motion of the electrons are
further confined in addition from that in the absence of the field.
The potential energy due to the electric field is given by

\begin{equation}
V_F(z) = \mp eFz
                                                   \label{efield}
\end{equation}

\noindent
where $-$ is for an electron and $+$ is for a hole. In this paper the
electronic charge is $-e$ ($<0$). As one increases $F$, carriers confined
initially by $V_w (z)$ would tunnel out of the well and therefore lowering
their potential energy.
If, however, the electric field is not extremely strong, the
quantum--confined states would have a long lifetime and can therefore be
considered as quasi--bound\cite{Bastard2}.

In modulation--doped SQW structures of GaAs/AlGaAs, electrons are
transferred from the impurities doped AlGaAs to the GaAs region so as to
lower the potential energy of electrons. As electrons transfer to the well
region, the conduction and valence band edges are modified from the flat
ones. This band--bending effect is to
be included in the calculation by solving the Schr\"{o}dinger equation with a
use of a self--consistent potential energy $V_{s.c.}(z)$,

\begin{equation}
V_{s.c.}(z) = V_H(z) + V_{xc}(z).
                                              \label{pot}
\end{equation}

\noindent
Here $V_H(z)$ is the Hartree interaction energy, which is the sum of two
contributions;
the electrostatic interaction of a given carrier with the other carriers
in the quantum well and that with ionized impurities remaining in the barrier
region. $V_{xc}(z)$ is a single particle version of the exchange--correlation
energy, which is not included in $V_H(z)$. In the Hartree
approximation one only considers $V_H(z)$ and neglects  $V_{xc}(z)$ in
$V_{s.c.}(z)$. The Hartree term $V_H(z)$ is a solution of a Poisson
equation. In this work, we consider the case that impurities are doped
symmetrically with a bulk density $N_d$ in each barrier region on both sides
of a SQW [see figure 1].
Therefore, the Poisson equation is written by\cite{Ando}

\begin{equation}
\frac{{\rm d}^2 V_H (z)}{{\rm d}z^2} = \mp \left \{ \begin{array}{ll}
\frac{4\pi e^2}{\epsilon} [n(z) - N_d] ,& L/2 < |z| < L/2 + d_b \\
\frac{4\pi e^2}{\epsilon} n(z) ,& {\rm otherwise}
\end{array}
\right.
                                                  \label{hart}
\end{equation}

\noindent
where the upper (lower) sign refers to electrons (holes).  In equation\
(\ref{hart}) $n(z)$ is the carrier density of the SQW and
$d_b$ is the width of the impurity--doped region in the barrier.
The total effective width of the SQW is $d = L + 2d_b$ [see figure 1].
The static dielectric constant of the quantum--well material $\epsilon$ is
assumed to be the same in both the barrier and well region.
The average two--dimensional ($2D$) carrier density $N_s$ of the well is
given by $N_s = 2N_dd_b$.
In this paper, we confine ourselves to the ground--state, zero--temperature
limit of the problem.
Hence the volume density distribution $n(z)$ is written by

\begin{equation}
n(z) = \sum_i N_{si}|\psi_i(z)|^2
                                                   \label{den}
\end{equation}

\noindent
where $N_{si}$ is the $2D$ density of carriers in the $i^{\rm th}$ subband
(i.e., $\sum_{i} N_{si} = N_s$).   If the electrons occupy only the ground
subband ($i = 1$), $N_s = N_{s1}$
[which is the case of weakly--doped samples in experiment], and we can
rewrite equation\ (\ref{den}) by

\begin{equation}
 n(z) = N_s |\psi_1(z)|^2.
						   \label{density}
\end{equation}

The Hartree approximation is, in general, known to overestimate the Coulomb
repulsive force of other electrons in the short range, and, in real system,
many--body effects such as the exchange--correlation interaction could become
important in the study of subband structures or intersubband resonance
absorption\cite{Dingle2}. We include the exchange--correlation effects in our
calculation by introducing a simplified local exchange--correlation potential
$V_{xc}(z) = V_{xc}[n(z)] \equiv \mu_{xc} [n(z)]$ within the local
density--functional approximation
(LDA)\cite{Hohenberg,Kohn,Sham}.
Here $\mu_{xc}$ is the
chemical potential of an inhomogeneous electron gas, which is obtained from
that of a homogeneous electron gas by replacing the uniform electron density
by the local electron density $n(z)$ of the inhomogenous system.
Several different forms of $V_{xc}(z)$ are proposed by a number of
people\cite{Katsnelson}. In our work, we use a simple form suggested by
Gunnarson and Lundqvist\cite{Gunnarson},

\begin{equation}
V_{xc}[n(z)]= - \frac {2}{\pi \alpha r_s} [1 + 0.0545r_s
\ln (1 + \frac {11.4} {r_s}) ] \frac {m^*e^4}{2 \epsilon^2 \hbar^2}
                                                   \label{exch}
\end{equation}

\noindent
where $\alpha = (4/{9\pi})^{1/3}$ and $r_s$ is defined in term of
$n(z)$ by $n(z) = [\frac {4\pi}{3} (a^{*}_{B} r_s)^{3} ]^{-1}$
with $a^{*}_{B} = {\epsilon \hbar^2}/{m^*e^2}$, the effective Bohr radius.
Therefore, the Schr\"{o}dinger equation for a single particle of the
potential energy $V_{s.c.}(z)$ given by equations\ (\ref{hart}) and
 \ (\ref{exch}) becomes the `so--called' Kohn--Sham equation.

In actual numerical calculation, one first needs to introduce impenetrable
potential barriers at the positions sufficiently far from both sides of the
SQW in such a way that the positions of the barriers do not affect the
solutions of the problem. In our calculation, we assume these barriers
of infinite height at $x= \pm L_B$ ($\approx \pm 2.5L$)\cite{Bloss} where
$-L/2$ and $L/2$ are the edges of the finite SQW [see figure 1].
We, then, write, within the centeral difference approximation\cite{Gerald},
the second--order differential equation, equation\ (\ref{schr}) by a
difference equation of unknown $\psi(z_i)$ for each point $z_i$,
a point of the $i^{\rm th}$ subdivision of the interval [$-L_B$, $L_B$].
Therefore, we replace the ordinary differential equation,
equation\ (\ref{schr}) by a finite difference equation;

\begin{equation}
\sum^n_{i=1} [\frac {\psi(z_{i+1}) - 2\psi(z_i) + \psi(z_{i-1})}
{\triangle ^2} + V_t(z_i) \psi(z_i)] = E \sum^n_{i=1} \psi(z_i).
                                                    \label{diff}
\end{equation}

\noindent
In equation\ (\ref{diff}), $\triangle$ is the grid spacing,
which is the length of
each subinterval and $n$ is the number of subintervals. The boundary
conditions require $\psi_o(z_o) = \psi_{n+1}(z_{n+1}) = 0$, where
$z_n = z_o + n\triangle$.   Writing equation\  (\ref{diff}) in a matrix form,
we
get an $n$ by $n$ tridiagonal matrix eigenvalue equation.
This eigenvalue equation is solved by diagonalizing the matrix.
In this work, we solve this problem self--consistently
with an accuracy such that the difference in the Hartree potential energies
of the $\it{i}^{\rm th}$ and ${(\it{i}+1)}^{\rm th}$ iterations is less
than $10^{-3}$meV. The matrix--size dependence of the eigenvalues
is given in figure 2, as an example.  It shows the electron subband energies
in the absence of the field within a single-particle scheme, that is,
$N_s = 0$; the solution of equation\ (\ref{schr}) with $V_t(z) = V_w(z)$.
Rapid convergence is observed in the eigenenergies as we increase the size
of the matrix in solving the equation\ (\ref{diff}).
We choose, in the rest of this paper, $500$ by $500$ as a relevant size of
the matrix in actual numerical calculations. In this case, the discrepancy
in eigenenergies obtained by the shooting and finite difference methods is
not greater than $0.1$meV.

\section*{3. Results and discussion}

\noindent
The subband energies of an electron and a hole in GaAs/AlGaAs SQW are
evaluated and the effects of carrier interaction and electric field are
examined.
The physical quantities used in obtaining numerical results are listed in
table I.   In this paper, we ignore the variation of the effective masses
in the well and barrier regions of the SQW. And we consider the case
that the bulk impurity density $N_d = 10^{18} {\rm /cm}^3$ in the rest of
the paper.

\subsection*{3.1. Self--Consistency Effects}

\noindent
The potential profiles of the modulation--doped SQW in the presence of an
external d. c. electric field $F = 30{\rm kV/cm}$ are shown in figure 3 for
three different surface carrier densities $N_s$. In the figure, the vertical
axis, especially the energy gap in the well is not drawn in scale.
In the well region, the bottom of the conduction band moves upwards as one
include the Hartree potential $V_H(z)$, i.e., the interaction
of a given electron with other electrons in the quantum well and also with
the ionized impurities doped in the barrier region.   Hence, the electron
subband energies are also expected to increase due to the Hartree correction.
In the valence band, however, the holes are more confined by the Hartree
potential than electrons in the conduction band.  This effect could cause a
reduction of the hole subband energies.
On the other hand, in modulation--doped SQW structures, there
occurs a band--bending effect due to the dipoles formed between the plus
(ionized donors) and the minus (electrons) charges and this effect would
modify the subband energies. The effects of doped ionized impurities on the
electron and hole subbands are different because of the opposite sign of the
electric charges of an electron and a hole.

Figure 4 shows electron subband energies $E_{en}$ as a function of the
carrier density $N_s$ for four different values of the electric field
strength $F$. The ground ($n = 1$) and first excited ($n = 2$) subband
bottom energies are displayed, respectively, in the figure 4(a) and (b).
The insets indicate that the electron subband energies are measured upwards
from the bottom of the conduction band of the well with $F = 0$ and
$N_s = 0$, which is taken the same as the midpoint of the conduction band
bottom of the well in the presence of the field within a single-particle
scheme; with $N_s = 0$ .
The $E_{en}$ decreases as one increases the magnitude of the electric field
$F$.
The reduction of the subband energies under the electric field is in
agreement with observations of quantum confined Stark effect\cite{Vina}.

The $N_s$ dependence of the subband energies is different in magnitude for
different subbands.
The subband energy $E_{e1}$ increases monotonically as $N_s$ increases.
[figure 4(a)]  However, the first excited subband energy $E_{e2}$ shows
different behavior.[figure 4(b)]  The $E_{e2}$ increases initially as $N_s$
increases.  For a given value of $F$, beyond a certain density $N_s^{\rm th}$,
$E_{e2}$ decreases until the subband $n = 2$ becomes quasi--bound to tunnel
out the barrier located at the right--hand side.  The ground subband energy
$E_{e1}$ increases more rapidly than that of the first excited subband
$E_{e2}$.
This different behavior is related to the effect of band--bending on the
barrier region.   Figure 5 shows the probability density $|\psi_{e}(z)|^2$
of the ground($n = 1$) and first excited ($n = 2$) electron subbands for
$F = 30$kV/cm and $N_s = 0.9 \times 10^{12}/{\rm cm}^2$ and
$1.7 \times 10^{12}/{\rm cm}^2$.
In the well region, as the surface carrier density $N_s$ increases, the
magnitudes of both $|\psi_e(z)|^2$ for both $n = 1$ and $2$ decrease.
On the other hand, the probability densities increase in the barrier
region as the $N_s$ increases.
The reduction of the $|\psi_e(z)|^2$ in the well region results from
the fact that the Hartree potential tends to locate the electrons near the
interfaces, raising the corresponding confinement energies.   However, the
effect of band bending reduces the effective width of the potential barrier
[see figure 3] and causes penetration of wave functions into the barrier
region.
The wave function penetration is more pronounced for the first excited
subband than the ground subband.
For the first excited subband, most of the electrons remain
near the edges of the well for $F \leq 50$kV/cm even in the absence of the
Hartree potential.   But, the electrons in the ground subband, if $F$ is not
extremely
large, such as, $F \leq 150$kV/cm, are well localized inside the quantum well
by the confining potential $V_t(z)$, in which the Hartree potential $V_H(z)$
is excluded.

Figure 6 shows the energy separation $E_{e2 - e1}$ of the ground and first
excited electron subbands as a function of $N_s$ for $F = $ 0, 30, 50, and
70 ${\rm kV/cm}$.
The separation of electron subband energies decreases as $N_s$ increases
because the effect of band bending is more pronounced in the first excited
subband than in the ground subband.
This result differs from the cases of Si inversion
layers\cite{Ksyi} or GaAs single heterostructures\cite{Stern}.
In the cases of Si inversion layers and GaAs heterostructures, subband
energy separations are observed to increase with increasing surface carrier
density.
This difference is conjectured to be related with the different boundary
conditions and the effective potentials $V_t(z)$. For the case of Si
inversion layers, the potential barrier at the interface of Si and SiO$_2$
is well approximated to be infinitely high so that the wave functions vanish
at the interface.
For the case of single heterojunction interface, the barrier height is
finite. The potential functions $V_t(z)$ in both cases are of triangular
shape and infinite number of subbands are expected to be bound.
However, in the SQWs, two heterojunction boundaries exist and the shape of
potential function is rectangular one with finite depth, hence, allowing only
a few subbands to be bound.

The ground hole subband energies $E_{h1}$ and first excited
hole subband energies $E_{h2}$ are shown, respectively, in figure 7(a) and
(b) as a function of $N_s$ for various electric fields $F$.
The hole subband energies are measured down from the midpoint of the valence
band top of the well region with $N_s = 0$, as is indicated in the
insets of figure 7(a) and (b).
{}From the carrier--density dependence of the potential energy profile of the
valence band [see figure 3], the confinement of the holes is expected to
increase as $N_s$ increases.
As one increases the carrier concentration of the SQW, the top of the valence
band moves up so that the holes are more tightly bound.    Hence, hole
subband energies decrease with increasing surface carrier density.
The electric--field dependences of $E_{h1}$ and $E_{h2}$ are opposite to each
other.  The first excited hole subband is less bound for higher electric
fields[see figure 7(b)].

Figure 8 shows the probability densities $|\psi_h(z)|^2$ of the
ground($n = 1$) and first excited($n = 2$) hole subbands in the presence of
an electric field $F = 30$kV/cm for two different surface carrier densities
$N_s = 0.9 \times 10^{12}/{\rm cm}^2$ and $ 1.7 \times 10^{12}/{\rm cm}^2$.
We observe that the wave functions are more confined near the center of
the well for higher $N_s$ and this confinement
is more pronounced for the case of the ground subband.
The first excited hole subband energies increase as $F$
increases as is shown in figure 7(b).
This peculiar behavior is caused by the fact
that the first excited wave function is more confined to the center of the
well as the electric field increases\cite{Matsuura}.
Figure 9 shows the energy separation of the ground and first excited hole
subband energies $E_{h2 - h1}$ as a function of $N_s$ for four different
values of the electric field strength $F$. We observe that the hole subband
energies separation increases as one increases either $N_s$ or $F$.
This observation differs from the case of electron subbands [see figure 6]
because of the different potential profiles[see figure 3].
In the zero--field case, the first
excited hole subband energy ($\approx 33$meV) is relatively small compare to
the valence band offset ($\approx 160$meV).   Therefore, the effect of the
Hartree potential on the first excited hole subband is much less than that
on the corresponding electron subband.

\subsection*{3.2. Many--Body Effects}

\noindent
We include the effects of exchange  and correlations in our calculation by
employing a simple local exchange--correlation potential energy $V_{xc}(z)$
given by equation\ (\ref{exch}).   Strickly speaking, the density--functional
formulation is good only for the
calculations of the ground state energy and electron density
distribution\cite{Hohenberg,Kohn,Sham}. However, it has been used, in a good
approximation, for the subband structure calculation\cite{Ksyi}.
The potential energy profiles of the conduction and valence bands of a SQW
are shown, respectively, in figure 11(a) and (b) in the
presence of an external d. c. electric field with (solid line) and without
(dotted line) the exchange--correlation potential. The exchange--correlation
potential slightly reduces the Coulomb repulsive force and electrons are
pushed further toward the interface\cite{Ando}.
However, in this work, we observe that
$V_{xc}(z)$ lowers the conduction and valence band profiles of the SQW by
negligiblely small amount.
The magnitude of the $V_{xc}(z)$ is so small, compared to other terms in
$V_t$, that its effect on the total potential energy profiles is quite small.
This result differs from the cases of Si inversion
and accumulation layers\cite{Ksyi,Stern}.

The subband energy separations of $E_{e2-e1}$  and $E_{h2-h1}$ are shown,
respectively, in figure 11 (a) and (b) as a function of
the surface carrier density $N_s$ with (solid line) and
without (dotted line) the exchange--correlation potential. The effect of the
exchange--correlation potential is not significant and is only to increase
subband energy separations slightly as is shown in the insets of figure
11(a) and (b).
This increase comes from the fact the ground subband energy is
lowered slightly more than that of the excited subbands, in agreement with
the cases of Si layers\cite{Ksyi}.
The negligible effect of the exchange--correlation potential on the subband
structure of the SQW system is conjectured to the relatively large
(kinetic) confinement energy of the system, as compared with the cases of the
single heterostructures\cite{Stern} or Si space charge layers\cite{Ksyi}.

\section*{4. Conclusions}

\noindent
We investigated the electronic properties of symmetrically--doped GaAs/AlGaAs
single quantum wells in the presence of an external d. c. electric field
perpendicular to the system and examined the effects of the field and
carrier interactions on the system.

We summarize the results as follows.
  First, we evaluated the subband energies and wave functions as a function
of surface carrier density.   A self--consistent calculation within a Hartree
approximation shows that, as the surface carrier density increases, the
ground electron subband energy increases monotonically,
but the first excited subband energy starts to decrease beyond a certain
density. However, the energy separation of the lowest two electron subbands
is reduced as the surface carrier density increases.
For the hole subbands, as the surface carrier density increases, individual
subband energy decreases, but their energy separation increases.
This different behavior in the carrier--density dependence of the electron
and hole subbands results from the different confinement effects for two
different charge carriers.
  Second, the exchange--correlation effect is examined within a local
density--functional approximation.
A small reduction of the subband energies and
an slight enhancement of the subband energy separation is observed.
However, the many--body effect is found not so significant.
This result is different from the cases of the Si inversion layers or
GaAs single heterostructures.

\acknowledgements
\noindent
This work was supported in part by the BSRI--94--234 program of the
Ministry of Education, Korea and in part by 1994 program of Korea Research
Foundation (project number 02-D-0141).

\newpage
\begin{center}
{\large \bf Figure Captions}\\
\end{center}

\noindent
{\bf Figure 1.} Schematic diagram of an idealized quantum well
structure of width $L$ and barrier height $V_o$ in the presence
of an external d. c. electric field $F$.
The wave functions are to vanish at $-L_B$ and $+L_B$ for
the convenience of a calculation.
The width of symmetrically doped region is indicated by $d_b$
in the barrier regions.\\

\noindent
{\bf Figure 2.} The variation of the electron subband energies as
a function of the size of matrix.   The results are the case of
undoped SQW in the absence of the electric field.  The width of
the well is 85\AA .\\

\noindent
{\bf Figure 3.} The potential energy profiles of modulation--doped
SQW in the presence of an electric field $F = 30$kV/cm for different
surface carrier densities $N_s$.  The width of the well is 85\AA.
The vertical axis, especially, the energy gap in the
well region is not drawn in scale.\\

\noindent
{\bf Figure 4.} The electron subband energies as a function of the surface
carrier density $N_s$ for various electric fields $F$.
The inset indicates that the energy is measured up from the midpoint of
the conduction band bottom in a single--particle scheme in the presence of
the fields.
(a). the ground electron subband.  (b). the first excited electron subband.\\

\noindent
{\bf Figure 5.} The probability density of the ground and first excited
electron subbands under an electric field $F = 30$kV/cm for two different
surface carrier concentrations $N_s$.   The width of the well is 85\AA . \\

\noindent
{\bf Figure 6.} The energy separation of the ground and first excited
electron subbands as a function of the surface carrier density $N_s$ for
four different values of electric field strength.\\

\noindent
{\bf Figure 7.} The hole subband energies as a function of the surface
carrier density $N_s$ for various electric fields $F$.
The inset indicates that the hole subband energies are measured down
from the midpoint of the valence band top in the well region in a
single--particle scheme in the presence of the fields.
(a). the ground hole subband.   (b). the first excited hole subband. \\

\noindent
{\bf Figure 8.} The probability density of the ground and first excited
hole subbands under an electric field $F = 30$kV/cm for two different
surface carrier concentrations $N_s$.   The width of the well is 85\AA .\\

\noindent
{\bf Figure 9.} The energy separation of the ground and first excited
hole subbands as a function of the surface carrier density $N_s$ for
four different values of electric field strength $F$.\\

\noindent
{\bf Figure 10.} The potential energy profiles with (soild line) and without
(dotted line) the exchange--correlation potential in the presence of an
electric field $F = 30$kV/cm for a constant surface carrier
density $N_s = 1.7\times 10^{12}/{\rm cm}^2$.
The width of the well is 85\AA .
(a). the bottom of the conduction band.   (b). the top of the valence band.\\

\noindent
{\bf Figure 11.} The energy separation of the lowest two subbands with
(solid line) and without (dotted line)
the exchange--correlation potential as a function of
the surface carrier density $N_s$ for an electric field $F = 30$kV/cm.
The solid and dotted lines are superposed nearly on top of each other.
(a). the electron subband energy separation.  (b). the hole subband energy
separation.\\

\begin{table}[h]
\caption{Physical quantities used in numerical calculation}
\begin{flushright}
($m_o$ : free electron mass)
\end{flushright}
\begin{tabular}{|c|c|}
Quantum Well Material & GaAs/AlGaAs SQW \\ \hline
Well width & 85\AA \\ \hline
Electron effective mass & $0.067m_o$ \\ \hline
Hole effective mass & $0.45m_o$ \\ \hline
Conduction band offset & 240meV \\ \hline
Valence band offset & 160meV \\ \hline
Background dielectric constant & 13.1 \\ \hline
Doped impurity density, $N_d$ & $1.0\times 10^{18}/{\rm cm}^3$ \\
\end{tabular}
\end{table}
\end{document}